\documentclass[onecolumn,preprintnumbers,amsmath,amssymb,showkeys]{revtex4-2}
\usepackage{graphicx}
\usepackage{amsmath}
\usepackage{amssymb}
\usepackage{bm}
\usepackage{orcidlink}
\usepackage{xcolor}
\usepackage{tabularx}

\begin{document}
\newcommand\ket[1]{\left| #1 \right>}
\newcommand\bra[1]{\left< #1\right|}
\newcommand\bracket [1]{\left( #1\right)}
\newcommand\sbracket [1]{\left[ #1\right]}

\newcommand{\orcidauthorA}{0000-0002-7619-3780} 
\newcommand{\orcidauthorB}{0000-0002-1668-8984} 
\newcommand{\orcidauthorC}{0000-0003-3377-1859}

\title{Mutual information and correlations  across topological phase transitions in topologically ordered graphene zigzag  nanoribbons}

\author{In-Hwan Lee\orcidlink{0000-0002-7619-3780}}
\thanks{These two authors contributed equally}
\author{Hoang-Anh Le\orcidlink{0000-0002-1668-8984}}
\thanks{These two authors contributed equally}
\author{S.-R. Eric Yang\orcidlink{0000-0003-3377-1859}}
\thanks{Corresponding author, eyang812@gmail.com}
\affiliation{Department of Physics, Korea  University, Seoul 02841, Korea
}




\begin{abstract}
Graphene zigzag nanoribbons, initially in a topologically ordered state, undergo a topological phase transition into crossover phases distinguished by quasi-topological order.
We computed mutual information for both the topologically ordered phase and its crossover phases, revealing the following results:
(i) In the topologically ordered phase, A-chirality carbon lines strongly entangle with B-chirality carbon lines on the opposite side of the zigzag ribbon. This entanglement persists but weakens in crossover phases.
(ii) The upper zigzag edge entangles with non-edge lines of different  chirality on the opposite side of the ribbon.
(iii) Entanglement increases as more carbon lines are grouped together, regardless of the lines' chirality.  No long-range entanglement was found in the symmetry-protected phase in the absence of disorder.

\end{abstract}

\keywords{topological order, mutual information, topological phase transition} 

\maketitle


\section{Introduction}

The topological order of graphene \cite{Nov, Zhang, Neto} zigzag nanoribbons \cite{Fujita,Yang, Brey, Brey2006} (ZGNRs)  manifests in the weak disorder regime through the presence of $e/2$  solitonic fractional  charges~\cite{Lei13, Wilczek03, Nakamura01, Barto1,Heeger} localized at opposite edges of the ribbon and spin--charge separation \cite{Yang2019, yang1,Yang2020}.
In this regime, interacting disordered ZGNRs  exhibit a universal value of topological entanglement entropy~\cite{Yang2021}, providing evidence for the existence of long-range entanglement and, consequently, topological order in these systems \cite{Kitaev11, Levin11}.
Additionally, in interacting disordered ZGNRs, the shape of the entanglement spectrum closely resembles the density of states of the edge states \cite{Yang2022}, aligning with the expectations for a topologically ordered system, as proposed by Li and Haldane \cite{Haldane191}.
Furthermore, a phase diagram of the topological order of ZGNRs predicts crossover phases {that exhibit non-universal values of topological entanglement entropy, each characterized by} distinct properties in the regimes of intermediate on-site repulsion and intermediate disorder strength \cite{Yang2023Phase}. In one of the crossover phases, crossover phase I, fractional charges and spin--charge separation are notably absent. However, charge transfer correlations persist between the zigzag edges.
In the other crossover phase crossover phase II, fractional charges are present, but no correlations are observed between the opposite zigzag edges.
The experimental verification of these theoretical predictions has an excellent chance of success since the fabrication of ZGNRs has achieved  atomic precision \cite{Cai2, Kolmer, Houtsma}.

In this study, we investigate how the entanglement pattern changes during the transition from either a crossover phase or a symmetry-protected phase to the topologically ordered phase.
Our investigation is motivated by several factors. Firstly, near the usual quantum-critical point, electrons cease to behave independently and, instead, become entangled in large numbers.   Entanglement patterns exhibit intriguing features near a quantum critical point, as previously described in  works by Sachdev et al. \cite{sachdev2011, sachdev2012}. This quantum phase transition shares some similarities with the smooth topological phase transitions from crossover  I/II phases to the topologically ordered phase. It is worthwhile to explore how the entanglement patterns evolve between these two phases.
Secondly, the transition from disorder-free interacting ZGNRs \cite{tan2014_band} (symmetry-protected phase) to their interacting disordered counterparts (topologically ordered phase) represents an abrupt topological phase transition. In other words, disorder acts as a singular potential that significantly alters the system's physics. It remains unclear how the entanglement pattern changes during this transition.

\begin{figure}[h!]
    \includegraphics[width=0.6 \textwidth]{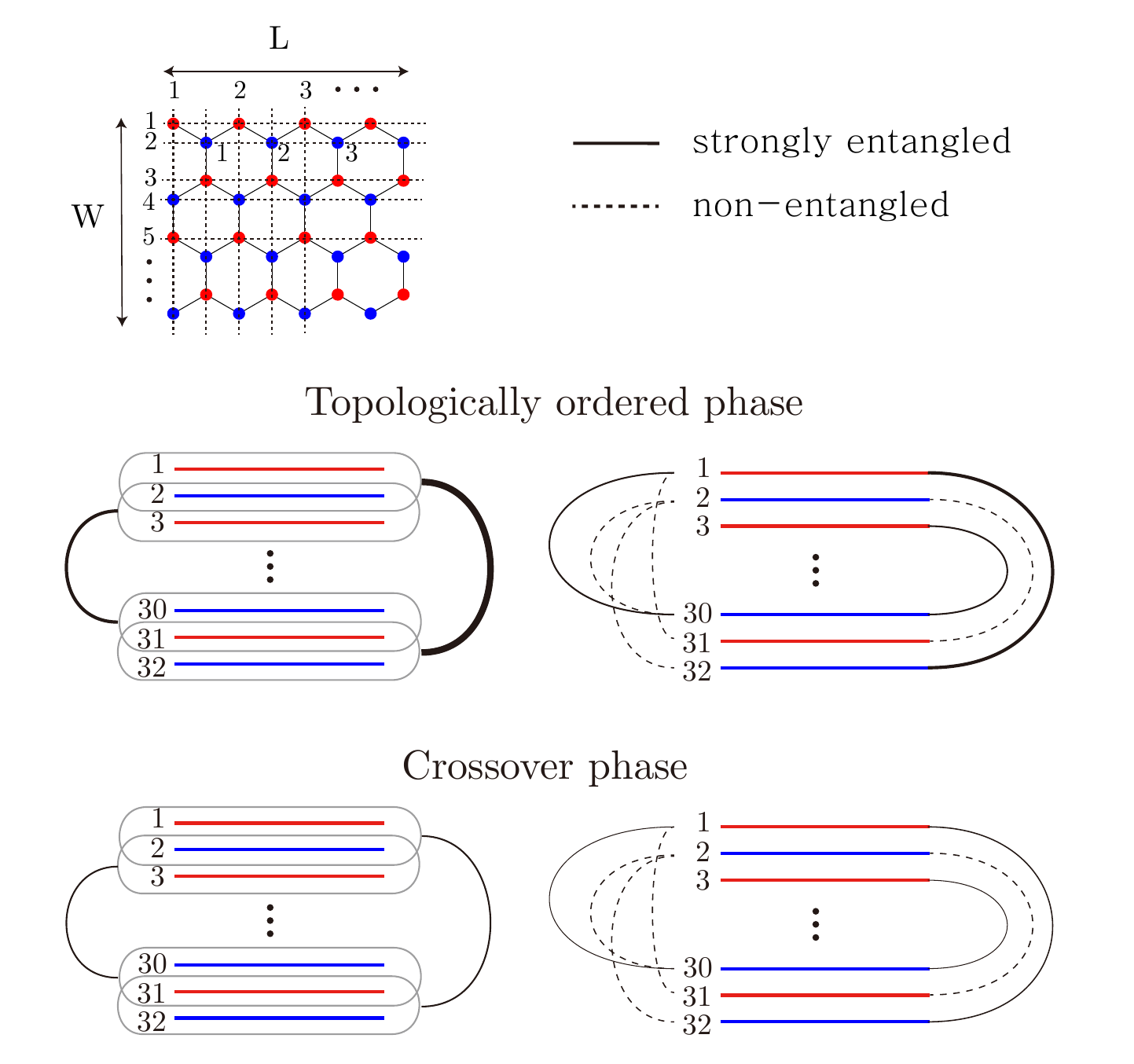}
    \caption{A zigzag ribbon consists of horizontal carbon lines with A  (red sites) and B (blue sites) chirality, as shown in the upper figure (disorder is not shown).   The upper zigzag edge consists of A-type carbon atoms, while the lower zigzag edge consists of B-type carbon atoms (mirror carbon lines have opposite chirality.) In the ground state, the upper edge sites have net spin-up while the lower edge 
    sites have net spin-down.
    Below is illustrated the entanglement of carbon lines represented by red and blue straight lines, in both topological order and crossover phases. The thickness of the solid curved lines represents the strength of the entanglement.  Dashed lines present weak entanglement between lines not connected by mirror symmetry.
 }
    \label{summary}
\end{figure}

To address these issues, we have computed the mutual information between different carbon lines in zigzag ribbons, and our findings are summarized in Figure \ref{summary}. In the topologically ordered phase, A-chirality carbon lines are strongly entangled with B-chirality carbon lines on the {\it opposite} side of the zigzag ribbon, but not with other lines. 
 In crossover phases, they are also entangled but to a lesser degree than in the topologically ordered phase. 
 Furthermore, the upper zigzag edge entangles with carbon lines of opposite chirality in the lower part of the ribbon, as indicated by the solid lines in Figure \ref{summary}. However, entanglement increases as more carbon lines are grouped together, regardless of the lines' chirality. Notably, we did not observe long-range entanglement in the symmetry-protected phase in the absence of disorder, but it increases suddenly with the addition of disorder.

\vspace{-6pt}

The rest of this paper is organized as follows. Section \ref{section2} introduces the Hamiltonian model of ZGNRs. Section \ref{section3} presents the method for calculating mutual information and provides relevant notations. Sections \ref{section4} and \ref{section5} are devoted to presenting our numerical results of mutual information in topologically ordered and crossover phases, respectively. Comments and conclusions drawn from our results will be given in Section \ref{section6}.

\section{Hamiltonian Model}\label{section2}
We employ the self-consistent Hartree--Fock approximation for our numerical calculations, as it has been proven to work effectively for graphene systems~\cite{Pisa1}, and its results are consistent with those obtained through density functional theory and  density matrix renormalization group methods~\cite{Yang2022, White1992, Schollwock2011}.
 
{Disordered ZGNRs can be described within the Mott--Anderson Hamiltonian \cite{Dob, Belitz, Byczuk,Altshuler} and may be approximated using the Hartree--Fock approximation}:
\begin{equation}
\begin{aligned}
H_{\text{MF}} = & -t\sum_{n.n.,\sigma} c^{\dag}_{m,\sigma}c_{n,\sigma} +\sum_{m,\sigma} V_m c_{m,\sigma}^{\dag}c_{m,\sigma} \\ 
&+U\sum_m[ n_{m,\uparrow}\langle n_{m,\downarrow}\rangle +n_{m,\downarrow}\langle n_{m,\uparrow}\rangle-\langle n_{m,\downarrow}\rangle\langle n_{m,\uparrow}\rangle]
+ \sum_m [s_{mx} \langle h_{mx}\rangle+s_{my} \langle h_{my}\rangle].
\label{MFhspin}
\end{aligned}
\end{equation}

Here, $c^\dagger_{m, \sigma}$, $n_{m,\sigma}$, and $s_{m x(y)}$ represent creation operators, occupation numbers, and spin operators at site $m$ with spin $\sigma$ ($\sigma = \uparrow (\downarrow)$ for spin-up (down)). The summation with 'n.n.' specifies nearest neighbor hopping.
The second term represents a short-range  disorder potential \cite{Lima2012}, with 10\% of the total ribbon's sites containing impurities. The magnitude of the potential at each site is drawn from a uniform distribution $[-\Gamma, \Gamma]$. 
{
There is nothing special about the value of 10\%, i.e.,  one can also use different values.  We examined ZGNRs with other values of disorder concentration $n_\text{imp}$  and disorder strength $\Gamma$. These results are presented in a previous paper \cite{Yang2019}.  Note that, in the self-consistent Born approximation, the disorder strength is not solely characterized by a single parameter, either $\Gamma$ or $n_\text{imp}$, but by their product   $\Gamma \sqrt{n_\text{imp}}$ \cite{Katanin}.  Additionally, we have investigated ZGNRs with off-diagonal disorder in Ref. \cite{Yang2020}. }
The third and fourth terms arise from on-site repulsion $U$. The fourth term accounts for self-consistent "magnetic fields", where $\langle h_{m x} \rangle = -2 U \langle s_{m x} \rangle$ and $\langle h_{m y} \rangle = -2 U \langle s_{m y} \rangle$. {Note that this term is zero  in the undoped  case because the ground state  spins are collinear. The decoupling of the onsite term used in the Hartree--Fock Hamiltonian is derived from
\begin{equation}
\langle n_{m, \uparrow} n_{m, \downarrow} \rangle = \langle c^\dagger_{m, \uparrow} c^\dagger_{m, \downarrow} c_{m, \downarrow} c_{m, \uparrow} \rangle \rightarrow \langle n_{m, \uparrow} \rangle \langle n_{m, \downarrow} \rangle - \langle s^+_m \rangle \langle s^-_m \rangle,
\end{equation}
where the spin rasing and lowering operators are
\begin{equation}
\begin{aligned}
s^+_m &= s_{mx} + i s_{my}, \quad s^-_m = s_{mx} - i s_{my}, \\
s^+_m &= c^\dagger_{m, \uparrow} c_{m, \downarrow}, \quad s^-_m = c^\dagger_{m, \downarrow} c_{m, \uparrow}, 
\end{aligned}
\end{equation}
with site spin operators 
\begin{equation}
\begin{aligned}
s_{mx} &= \frac{1}{2} \left[ c^\dagger_{m, \uparrow} c_{m, \downarrow} + c^\dagger_{m, \downarrow} c_{m, \uparrow}  \right], \\
s_{my} &= -\frac{i}{2} \left[ c^\dagger_{m, \uparrow} c_{m, \downarrow} - c^\dagger_{m, \downarrow} c_{m, \uparrow}  \right], \\
s_{mz} &= \frac{1}{2} \left[ c^\dagger_{m, \uparrow} c_{m, \uparrow}- c^\dagger_{m, \downarrow} c_{m, \downarrow}  \right].
\end{aligned}
\end{equation}
(A detailed derivation of this term using Hartree--Fock approximation can be found in chapter 6 of Ref.~\cite{Yang}.)}
The relevant parameters in our study include on-site repulsion, disorder strength, and the number of doped electrons, denoted as $(U, \Gamma, \delta N)$.

The total number of carbon lines, denoted as $W$, and the total number of sites on each line, represented by $L$, are assigned as the width and length of the ZGNR, respectively. Therefore, the total number of carbon sites is given by $N_s = L \times W$. A Hartree--Fock single-particle state, denoted as $\ket{k}$ ($k=1,2,\ldots,2N_s$), can be expressed as a linear combination of site states $\ket{m,\sigma}$, with the coefficients $A_{k, m, \sigma}$ representing the self-consistent Hartree--Fock eigenvectors. In the language of second quantization, this relation can be represented as
\begin{equation}
a_k=\sum_{m,\sigma} A_{k,m,\sigma}c_{m,\sigma}.
\label{asumc}
\end{equation}

{The eigenstates of mean-field Hamiltonian in Equation (\ref{MFhspin}) are given by
\begin{equation}
\psi_\alpha = 
\begin{pmatrix}
A^\alpha_{1 \uparrow} & A^\alpha_{2 \uparrow} & \cdots A^\alpha_{N_s \uparrow} & A^\alpha_{1 \downarrow} & A^\alpha_{2 \downarrow} & \cdots A^\alpha_{N_s \downarrow}
\end{pmatrix}^T.
\end{equation}

With such representation of the eigenstates, the expectation values of site spins and site occupation numbers can be written as follows:
\begin{equation}
\begin{aligned}
&\langle s_{m x} \rangle=\frac{1}{2}\sum_{\alpha}{n_{\alpha}(A_{m\uparrow}^{\alpha*}A_{m\downarrow}^{\alpha}+A_{m\downarrow}^{\alpha*}A_{m\uparrow}^{\alpha})},\\
&\langle s_{m y} \rangle=\frac{i}{2}\sum_{\alpha}{n_{\alpha}(-A_{m\uparrow}^{\alpha*}A_{m\downarrow}^{\alpha}+A_{m\downarrow}^{\alpha*}A_{m\uparrow}^{\alpha})}, 
\end{aligned} 
\label{aspin}
\end{equation}
and
\begin{equation}
\begin{aligned}
&  \langle n_{m \uparrow} \rangle=\sum_{\alpha}{n_\alpha}(A_{m\uparrow}^{\alpha*}A_{m\uparrow}^{\alpha}), \\
& \langle n_{m\downarrow} \rangle=\sum_{\alpha}{n_\alpha}(A_{m\downarrow}^{\alpha*}A_{m\downarrow}^{\alpha}),\\
\end{aligned} 
 \label{nud}
\end{equation}
where $n_\alpha = 1 (0)$ for occupied (unoccupied) Hartree--Fock states.}

For the Hartree--Fock calculations, the input consists of an antiferromagnetic initial state for the disorder-free case at half-filling ($\Gamma = 0, \delta N = 0$), and a paramagnetic initial state otherwise. The paramagnetic initial state exhibits a small imbalance between spin-up and spin-down occupations at every site, resulting in HF eigenvectors with spin splitting~\cite{Dob, Yang2019}. {The paramagnetic initial state produces a state with fractional charges.}
The dimension of the Hartree--Fock matrix can scale up to 16,000, which corresponds to the longest ribbon considered in this study. The Hartree--Fock eigenstates and eigenenergies are computed self-consistently through 20 iterations to achieve convergence. We employed GPUs to expedite the solution of the Hartree--Fock matrix. Additionally, due to the extensive nature of the calculations involving numerous disorder realizations, GPU computations in this paper were conducted on a supercomputer. {Note that the ground state is doubly degenerate, with sites reversed. In undoped ribbons, site spins are collinear, while in doped ribbons, they are non-collinear.}

\section{Mutual Information}\label{section3}

\begin{figure}[h!]
\includegraphics[width=0.8 \textwidth]{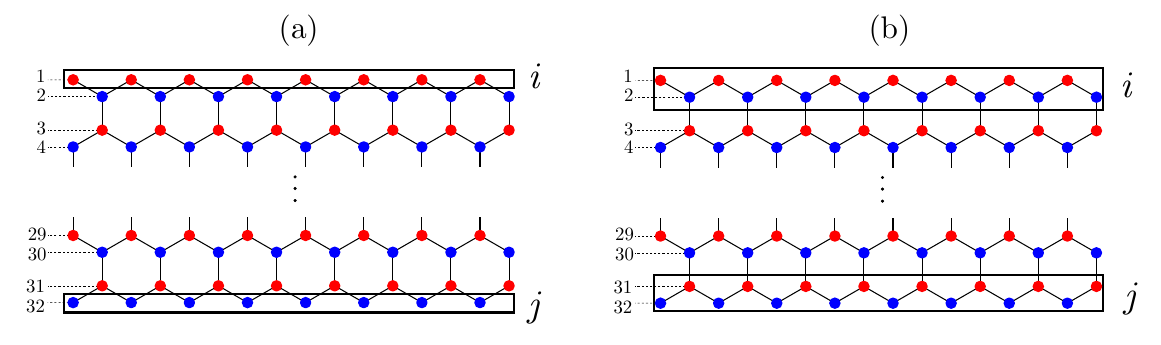}
\caption{Schematic figure for mutual information between carbon lines in ZGNR. The dashed horizontal lines specify the order of the carbon lines, and red (blue) dots represent A (B) type carbon atoms according to chiral symmetry. The ribbon's width is fixed at $W = 32$, as explained in the main text. (\textbf{a}) Mutual information between two zigzag edges (or between lines in the pair) with $i=1$ and $j=32$. (\textbf{b}) Mutual information between the top two lines and the bottom two lines with $i=1 \cup 2$ and $j= 31 \cup 32$.}
\label{M_region}
\end{figure}

In topologically ordered ZGNRs, two peaks in the probability density of fractional charges are located on the zigzag edge with nearly the same horizontal coordinates, as previously reported \cite{Yang2019}. Additionally, occupation numbers are correlated between pairs of carbon lines. These properties have motivated us to examine mutual information between carbon lines in a ZGNR. Mutual information is defined as follows:
\begin{equation}
M(i, j) = S(i) + S(j) - S(i,j),
\end{equation}
where $i$ and $j$ can represent single or multiple carbon lines ({domains}). Here, $S(i)$ and $S(i,j)$ denote the entanglement entropy of spin-up electrons within their respective areas.
{The entanglement entropy of domain $i$ is defined as the von-Neumann entropy \cite{Bal}
\begin{equation}
S (i) = - \text{Tr} (\rho_i \ln \rho_i),
\end{equation}
where $\rho_i$ is the reduced density matrix.
Within the Hartree--Fock approximation, it is comparatively convenient to adopt Peschel's method and compute the correlation matrix between sites $m$ and $n$ restricted within domain $i$ \cite{Peschel119}
\begin{equation}
    \mathbf{C}_{mn} = \bra{\Psi} c^\dagger_{m,\uparrow} c_{n,\uparrow} \ket{\Psi},
\end{equation}
where
 $\ket{\Psi}$ is one of the degenerate ground states. The reduced density matrix and spin correlation matrix $\mathbf{C}_{mn}$ are related by the following identity:
\begin{equation}
    \mathbf{C}_{mn} = \text{Tr} (c_{m,\uparrow}^\dagger c_{n,\uparrow} \rho_i).
\end{equation}

Thus, the entanglement entropy is equal to the sum of binary entropies \cite{Amico} of each eigenvalue $\lambda_i$ 
\begin{equation}
    S(A) = -\sum_p H(\lambda_p) = -\sum_p \left[ \lambda_p \ln \lambda_p + (1-\lambda_p) \ln (1 - \lambda_p) \right].
\end{equation}

These measures can also be applied to spin-down electrons.}
Mutual information, $M(i, j)$, quantifies the total amount of correlations between {domains} $i$ and $j$, and it obeys boundary laws \cite{wolf2008}. The notation for $S(i,j)$ is based on the following convention:
For the purposes of our analysis, we maintain a fixed ribbon width of $W = 32$. We employ the following notation: $(i, j) = (1, 32)$, which constitutes the first pair, comprising a line and its mirror counterpart in the opposite half of the ribbon. Continuing in this manner, $(i, j) = (2, 31)$ corresponds to the second pair, signifying that it is the closest pair to the outermost one, while $(i, j) = (3, 30)$ forms the third pair, and so forth. It is important to note that the two lines in these pairs are symmetrically positioned along the width of the ribbon. A visual representation of this notation is provided in Figure \ref{M_region}.

\section{Results for Topologically Ordered Phase}\label{section4}

\begin{figure}[h!]
    \includegraphics[width=1 \textwidth]{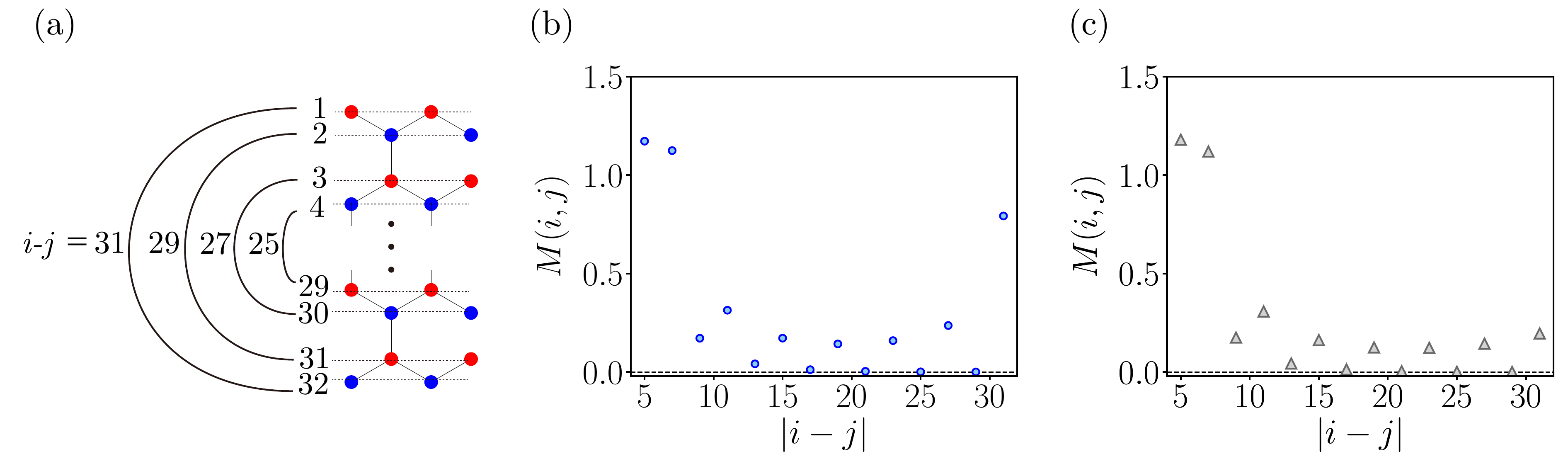}
    \caption{
    (\textbf{a}) Distances between symmetrically positioned lines are illustrated graphically.
    Numerical results for mutual information between symmetrically positioned lines (along the ribbon's width) as a function of their distance in the topologically ordered phase (\textbf{b}) with $(U, \Gamma, \delta N, N_D) = (t, 0.1t, 0, 20)$ and in  the  symmetry-protected phase (\textbf{c}) with $(U, \Gamma, \delta N, N_D) = (t, 0, 0, 1)$.}
    \label{M_TotLine}
\end{figure}

Let us delve into the entanglement patterns of the topologically ordered phase. To accurately quantify  the long-range entanglement, it is essential that the two lines under consideration be separated by a distance  greater than the correlation length. Figure \ref{M_TotLine} displays the dependence of the average mutual information between two symmetrically positioned lines on their distance, computed for ribbons with a width of $W = 32$. The mutual information values for the outermost pair differ significantly between the topologically ordered phase and symmetry-protected phase, implying the presence of long-range entanglement in the topologically ordered phase. Otherwise, the trends of mutual information in both the disorder-free and weakly disordered cases exhibit similarities, suggesting a comparable correlation length in both scenarios.
As we approach the innermost pair, the mutual information values abruptly increase, indicating that the distance between two lines decreases to less than the correlation length.
These results imply that the two lines, denoted as $i$ and $j$, should be separated by at least $10$ units of the lattice constant to ensure accurate mutual information calculations. Consequently, all mutual information calculations hereafter will be performed with a fixed ribbon width of $W = 32$, focusing primarily on the first few outermost pairs.  

\begin{figure}[h!]
    \includegraphics[width=0.9 \textwidth]{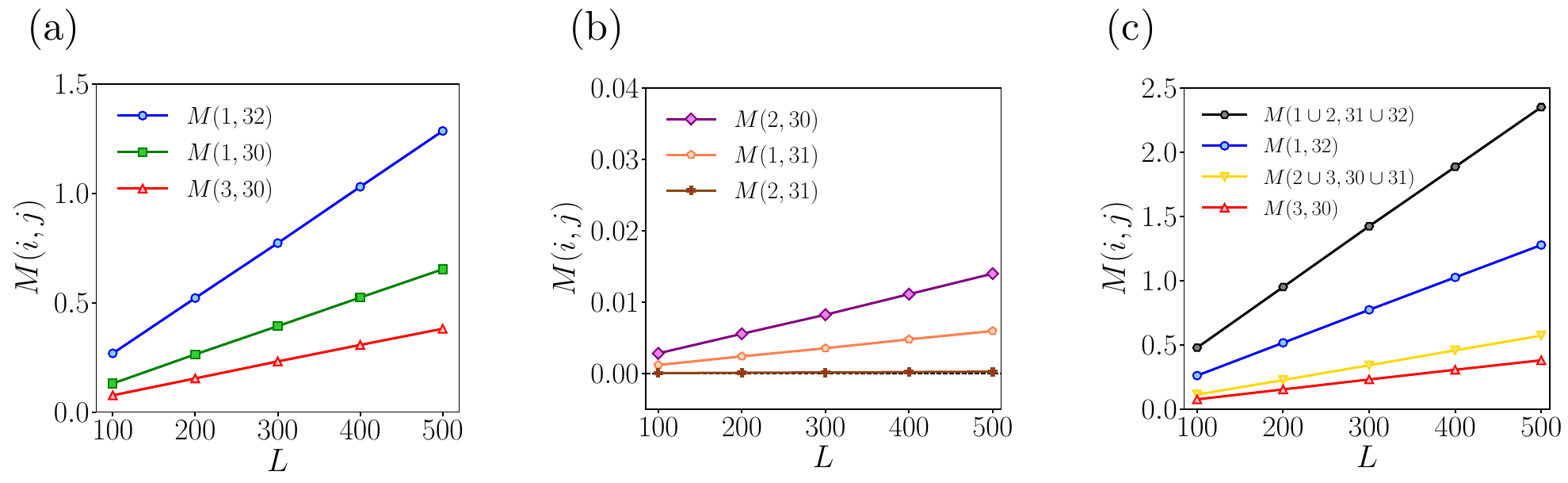}
    \caption{Numerical results for the dependence of mutual information on the length of ribbon with $(U, \Gamma, \delta N, N_D) = (t, 0.1t, 0, 20)$.
    The ribbon is in the topologically ordered phase.  (\textbf{a}) $M(i,j)$ for $(i,j)=(1,32),(3,30),(1,30)$. (\textbf{b}) $M(i, j)$ for $(i, j)=(2,30),(1,31),(2,31)$.  Due to the fact that the corresponding values of mutual information are nearly zero,  even numbered carbon lines in the upper half of ZGNR barely {entangle} with {their symmetric lines} in the lower half of the ribbon.   
    (\textbf{c}) In the double-line configuration, we depict $M(i,j)$ for $(i,j) = (1 \cup 2, 31 \cup 32)$ and $(2 \cup 3, 30 \cup 31)$ by using the values of $M(i,j)$ for $(1, 32)$ and $(3, 30)$ for comparison.  Note that the scales of $y$-axis differ among (\textbf{a}--\textbf{c}).} 
    \label{M_U1}
\end{figure}

There are several other noteworthy properties. Additional properties of mutual information  for a weakly disordered ZGNR with $(U, \Gamma,\delta N) = (t, 0.1t,0)$ are presented in Figure \ref{M_U1}a,b. 
Firstly, mutual information follows the area law, showing a linear scaling with the length of the ribbon. This behavior arises from the fact that the number of mixed chiral states, including those carrying fractional charges, is proportional to the length of the ribbon \cite{Yang2020}.  
 {We also observe that, when both $i$ and $j$ consist of two lines, the mutual information follows the same linear scaling, as depicted in Figure}~\ref{M_U1}c. The entanglement entropies $S(i)$ in the double-line configuration are smaller than those in the single-line configuration, as shown in Table \ref{tab1}. However, intriguingly, the mutual information in the double-line configuration is larger than that observed when $i$ and $j$ each consist of a single line. Specifically, the mutual information for $(i,j)=(1 \cup 2, 31 \cup 32)$ is twice as large as $(i,j)=(1,32)$, while the mutual information for $(i,j)=(2 \cup 3, 30 \cup 31)$ is one and a half times bigger than $(i,j)=(3, 30)$ (the third  pair).

\begin{table}[h!]
\caption{Numerical values of entanglement entropy, denoted as $S(i)$ and $S(j)$, were computed for $(U, \Gamma) = (t, 0.1t)$ using a single disorder realization ($N_D = 1$). In the case of double lines, values of entanglement entropy are smaller than those in the single-line case, whereas the mutual information $M(i,j)$ in the double-line case is larger.}
\label{tab1}
\newcolumntype{C}{>{\centering\arraybackslash}X}
\begin{tabularx}{\textwidth}{|C|CCCC|}
\hline
\textbf{}	& $S(i)$	& $S(j)$ & $S(i,j)$& $M(i,j)$\\
\hline 
Double lines: $ i = 1 \cup 2$, $j = 31 \cup 32$		& 40.19 	& 40.13  & 79.89 &  0.4299\\
\hline 
Single line: \ $ i = 1$, $j =  32$		& 59.01	& 58.93  & 117.7 &  0.2297\\
\hline
\end{tabularx}
\end{table}

\begin{figure}[h!]
\includegraphics[width=0.4 \textwidth]{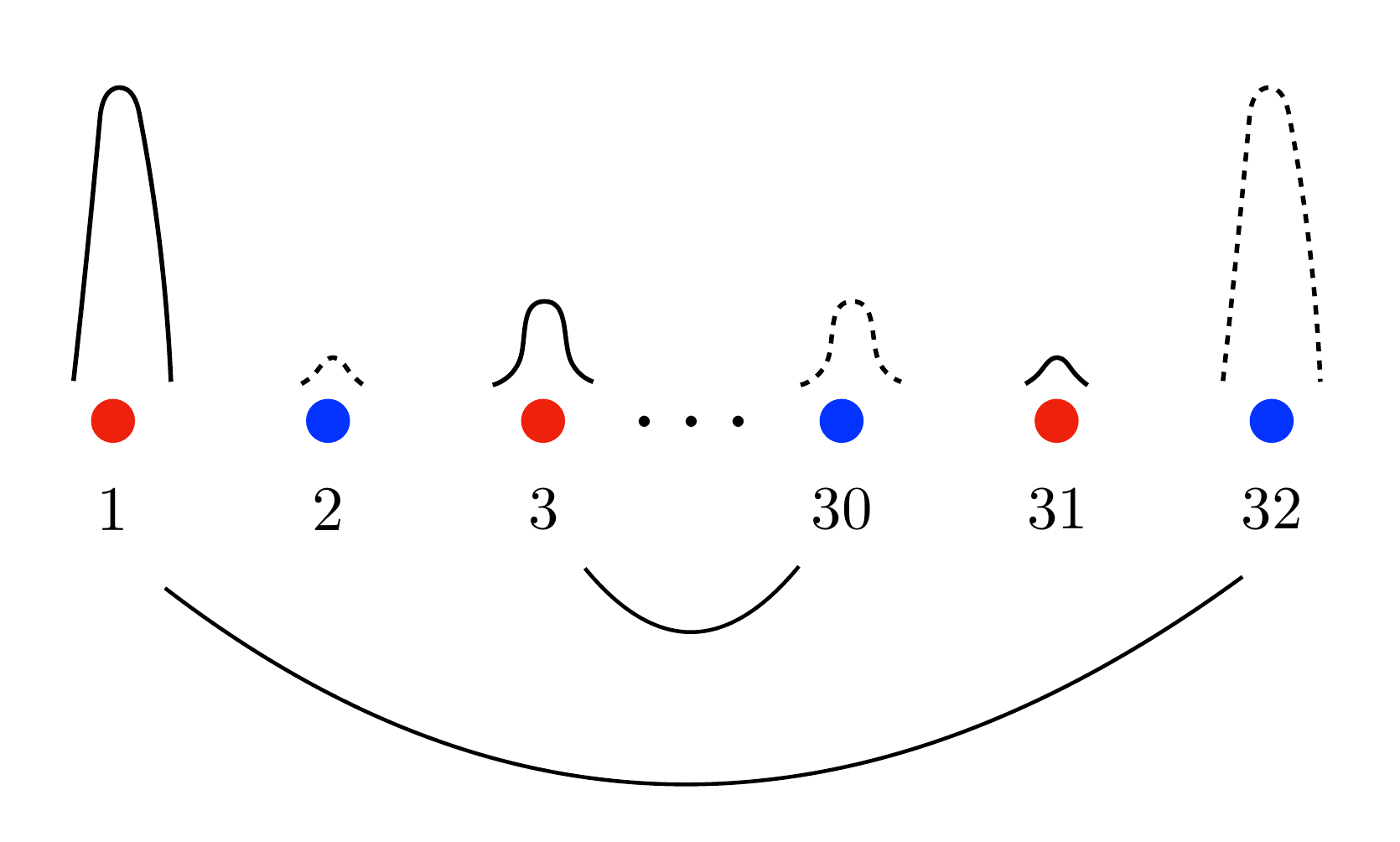}
\caption{ {Illustration of} 
the probability distribution of a mixed chiral state in the absence of disorder.  The horizontal carbon lines run perpendicular to the page. Solid   (dashed) line represents probability density distributed on $A$ ($B$)  sites. Odd-numbered pairs of lines are indicated.} 
\label{summary1}
\end{figure}

We find that mutual information vanishes for the even-numbered second, fourth, sixth, and so on, outermost pairs, while correlations are observed among  the  odd-numbered first, third, fifth, and so forth counterparts. As the  pair number increases, the mutual information between lines in the pair decreases. Additionally, there is a non-zero correlation between one edge and the 30th carbon line, and its mutual information ($M(1, 30)$) is even higher than that of the third  pair ($M(3, 30)$). 
These properties can be explained by the mixed chiral states depicted in Figure~\ref{summary1}. These states consist of two nonlocal parts residing on the edges of the $A$ and  $B$ sublattices, with their probability densities peaking at these edges and rapidly decaying within the ribbon. Consequently, the density distributed on the first line is greater than that on the third line, so we can deduce that $M(1, 30) > M (3, 30)$.  Note also that odd-numbered pairs exhibit substantial probability density on both their $A$ and $B$ carbon lines, while even-numbered pairs exhibit diminishing probability density  on their $A$ and $B$ carbon lines.

\section{Results for Crossover Phases}\label{section5}

We discuss how the behavior of mutual information is related to the topological transitions into crossover I and crossover II phases of interacting disordered ZGNR. Before we proceed, we briefly mention that at the critical point $\Gamma = 0$, where the symmetry-protected phase transitions to the topological order phase, mutual information exhibits a singularity, as seen in Figure~\ref{M_U,gdiff}b.
In the limit as $\Gamma/U \rightarrow 0$ with a ribbon length of $L = 300$, the mutual information between two opposite edges is $M(1,32) \approx  0.78 $ for $(U, \Gamma) = (t, 0.1t)$ and $M(1,32) \approx 0.20$ for $(U, \Gamma) = (t, 0)$.

\begin{figure}[h!]
\includegraphics[width=0.6 \textwidth]{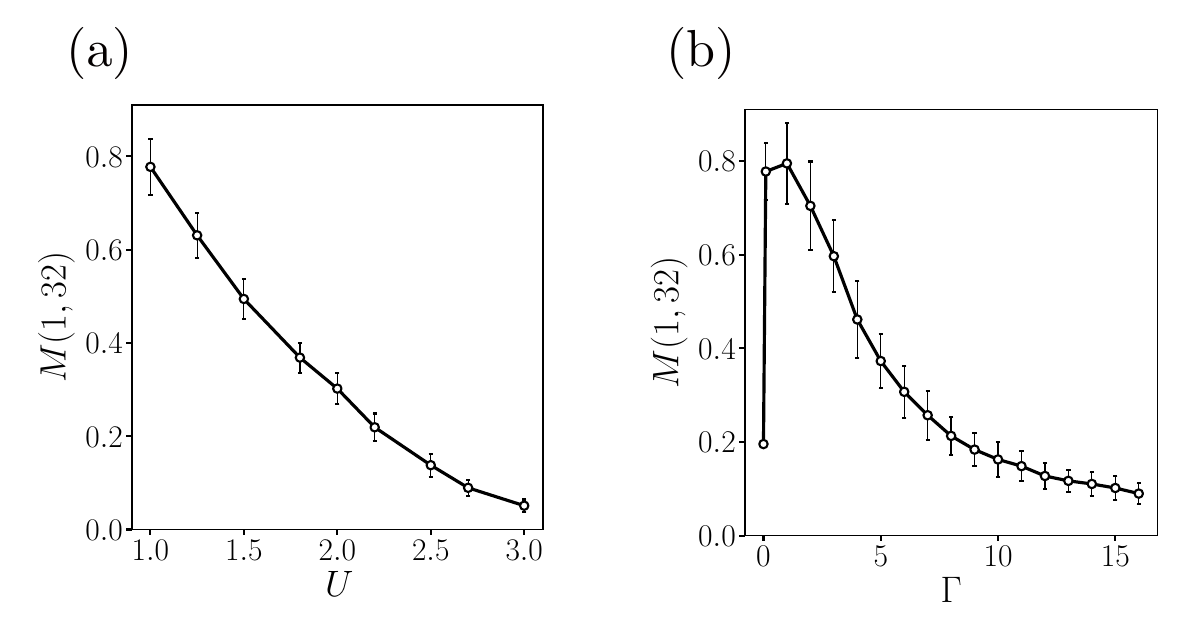}
\caption{ {(\textbf{a})} 
{Mutual information} between zigzag edges $M(1,32)$ with respect to on-site repulsion $U$. The other parameters are $(\Gamma, \delta N, N_D) = (0.1t, 0, 50)$. The ribbon is in the undoped {crossover I} phase.
(\textbf{b}) {Mutual information} between zigzag edges $M(1,32)$ with respect to disorder strength $\Gamma$. The other parameters are $(U, \delta N, N_D) = (t, 0, 50)$. The ribbon is in the undoped {crossover II} phase.
The error bars represent the standard deviation of each data point. A reduction in the mean value is observed with the increase in on-site repulsion and disorder strength, leading to a decrease in the size of the error bars. The ribbon size used in calculations is $(L, W) = (300, 32)$.}
\label{M_U,gdiff}
\end{figure}

In the crossover I and crossover II phases, mutual information  {decreases} 
upon increasing the strength of disorder and on-site repulsion, as seen in Figure~\ref{M_U,gdiff}. 
By increasing the strength of the disorder and on-site repulsion, the interacting disorder ZGNR undergoes a phase transition into a crossover phase and eventually a non-topologically ordered phase. Furthermore, the strength of the disorder affects mutual information in a manner similar to the impact of on-site repulsion, not only for the first pair but also for other lines: mutual information decreases as the disorder strength $\Gamma$ increases.

It is noteworthy that the behavior of mutual information mirrors that of topological entanglement entropy. Topological entanglement entropy experiences an abrupt increase from zero to a universal value when transitioning from the symmetry-protected phase to the topologically ordered phase \cite{Yang2021}. However, as the phase of the interacting disordered ZGNR undergoes a crossover from topologically ordered phase to non-topologically ordered phase \cite{Yang2023Phase}, topological entanglement entropy gradually decreases to zero and exhibits increasing variance.
Additionally, in the topologically ordered phase, the mean value of topological entanglement entropy slightly decreases (with unchanged variance) when both on-site repulsion and disorder increase \cite{Yang2023Phase}. A similar behavior is observed with mutual information, as shown in Figure~\ref{M_U,gdiff}: mutual information and topological entanglement entropy for $(U,\Gamma)=(2t,0.1t)$ and $(U,\Gamma)=(t,2t)$ are smaller than those for $(U,\Gamma)=(t,0.1t)$, and all these values correspond to the topologically ordered phase.

\begin{figure}[h!]
\includegraphics[width=0.78 \textwidth]{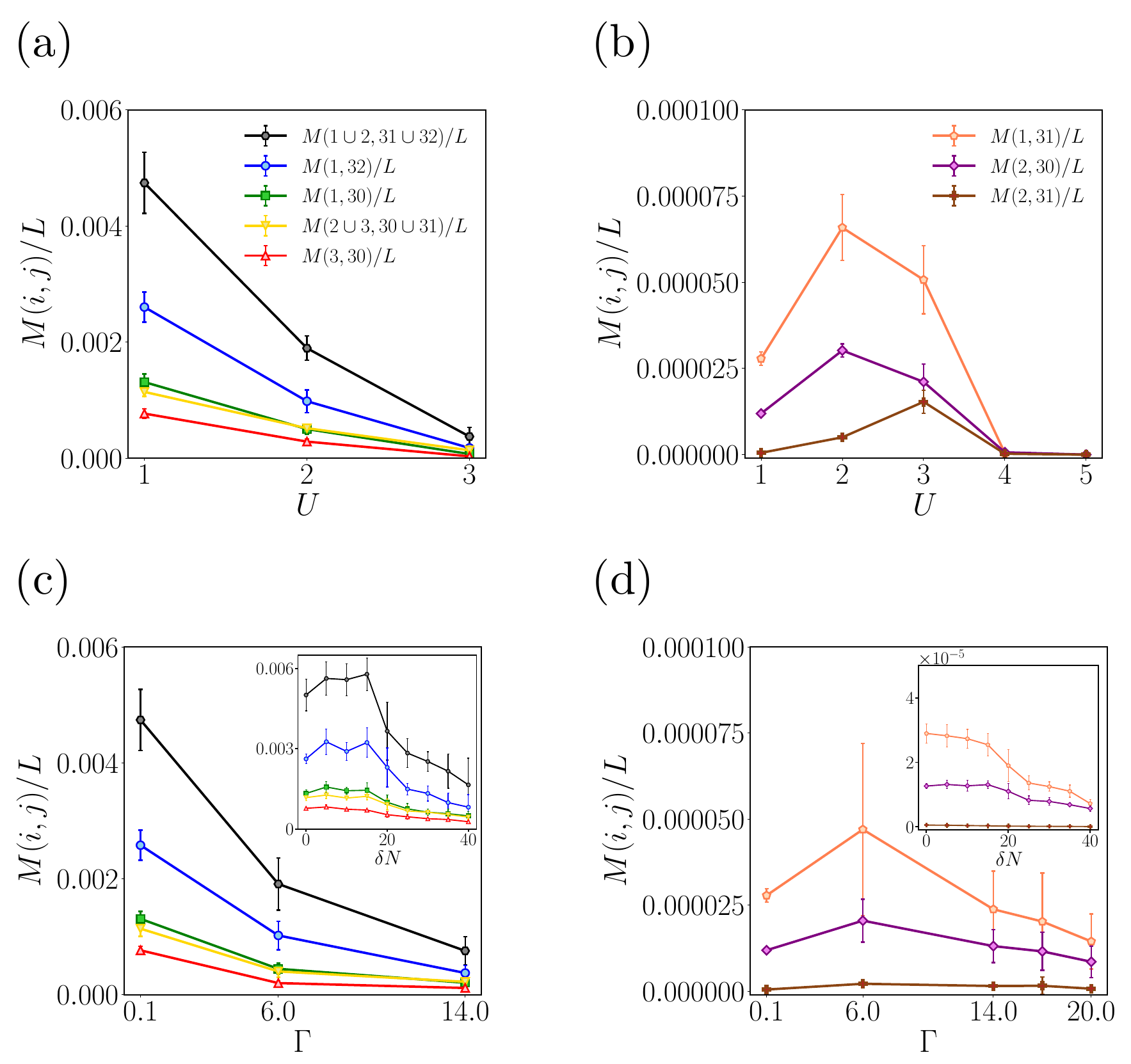}
\caption{ {Plot of the mean} 
value of mutual information divided by the ribbon length $L$ as a function of on-site repulsion. The parameters are $(\Gamma,\delta N)=(0.1t,0)$ for all cases, and $N_D$ was sampled more  than 20 times for $U = t, 2t$, 50 times for $U = 3t$, and 10 times for $U = 4t, 5t$.  The pairs of single carbon lines considered in the mutual information are $(i, j)=(1,32),(3,30),(1,30),(1\cup 2,31 \cup 32),(2\cup 3,30 \cup 31)$ in (\textbf{a}) and $(i, j)=(2,30),(1,31)$ in (\textbf{b}).  $M(1\cup 2,31 \cup 32)$ is twice as big as $M(1,32)$.  The trend of mutual information behaves similarly as disorder strength $\Gamma$ increases. 
{In (\textbf{c},\textbf{d}), the parameters are set to $(U,\delta N)=(0.1t,0)$ for all cases  {(color  lines  are  as  shown  in     figures  (\textbf{a})  and  (\textbf{b})).} The value of $N_D$ was increased to more than 20~times for $\Gamma=0.1t$ and $6t$, 50 times for $\Gamma=14t$, and 10 times for $\Gamma=17t$ and $20t$. Insets in (\textbf{c},\textbf{d}) depict $M(i,j)/L$ for various doping levels, $\delta N$, with $(U,\Gamma) = (1t, 0.1t)$ in a ribbon of size $(L,W) = (100,32)$. The doping concentrations corresponding to $\delta N$ in the figures range from $0$ to $0.0125$.}
}
\label{M_U123}
\end{figure}

We provide a detailed description of how carbon lines become entangled during the transition in the crossover phases, as depicted in  Figure \ref{M_U123}. As the topologically ordered phase transitions into crossover with an increase in on-site repulsion ($U$), the mutual information of the first pair $M(1, 32)$, the third pair $M(3, 30)$, and $M(1, 30)$ monotonically decreases (see  Figure \ref{M_U123}a). The mutual information for double-line configurations also follows the same decreasing trend (phase transition into the crossover II phase shows similar results to those of crossover I). This decline in mutual information with increasing $U$ is associated with the decrease in the number of well-localized fractional charges at the edge. Mutual information values associated with pairs with {the} next outmost lines, such as $M(2, 31)$, $M(2, 30)$, and  $M(1, 31)$, also decrease for  sufficiently large $U>3$, as shown in Figure \ref{M_U123}b. These values in the crossover phase show larger fluctuations compared to those shown in Figure \ref{M_U123}a  (note that, when $U \gtrsim 4$, the ribbon is not in the crossover phase). In addition, the phase of interacting disordered ZGNRs can transit to the crossover II phase upon increasing {the} number of doped electrons $\delta N$. The insets in Figure~\ref{M_U123}c,d show that mutual information has a decreasing trend with higher doping concentration.

Note that both mutual information and topological entanglement entropy are numerical values. Thus, they alone cannot distinguish between the crossover I and crossover II phases, as seen in  Figure \ref{M_U123}. To differentiate between these phases, a comprehensive analysis of the presence of fractional charges and edge charge correlations, as outlined in \cite{Yang2023Phase}, is required.

\section{Discussion}\label{section6}

Mutual information provides a quantitative measure of the entanglement between carbon lines. We calculated mutual information for both the topologically ordered phase and crossover phases. The entanglement correlations between carbon lines in an interacting disordered ZGNR are summarized in Figure \ref{summary} as follows:
(i) Mutual information scales linearly with the ribbon's length in all cases.
(ii) In both the topologically ordered and crossover phases, correlations exist among the first, third, fifth, and so on, pairs, while they are absent in the second, fourth, sixth, and so on, pairs.  These properties can be well explained by mixed chiral states.  Additionally, correlations exist between a zigzag edge and the even-numbered lines in the opposite half  of the ribbon, which have different chirality. {Our numerical results also show that all the non-vanishing correlations  weaken as on-site repulsion or disorder strength increases.}
(iii) The {mutual information among the outermost four lines (two from the upper and the other two from lower parts of the ribbon)} appears to be approximately twice as large as that among outermost pairs (first pair). However, it is worth noting that mutual information alone is not sufficient to distinguish between the crossover I and crossover II phases.
(iv) We also found that entanglement increases as more carbon lines are grouped together, regardless of the lines' chirality.
(v) We did not observe long-range entanglement in the symmetry-protected phase in the absence of disorder, but it increased suddenly with the addition of disorder.

It would be interesting to analyze the entanglement patterns of topologically ordered ZGNRs using string theory, as has been achieved near quantum critical points \cite{sachdev2011, sachdev2012}.  However, it is worth noting that the physics of crossover phases in topological phase transitions may not be identical to that near quantum critical points.

\section*{author contributions}
I.-H.L. and H.-A.L. performed the Hartree--Fock calculations.  S.-R.E.Y. conceived the project and supervised the study. All authors contributed to the writing of the manuscript.  {All authors have read and agreed to the published version of the manuscript.}

\bibliography{reference.bib}

\end{document}